\documentclass{article}
\usepackage[T1]{fontenc} 
\usepackage[utf8]{inputenc} 
\usepackage{ismir,amsmath,cite,url}
\usepackage{amssymb}
\usepackage{graphicx}
\usepackage{color}
\usepackage{booktabs}
\usepackage[bookmarks=false]{hyperref}

\usepackage[dvipsnames]{xcolor}
\newcommand\myshade{70}
\colorlet{mywholecolor}{MidnightBlue}
\hypersetup{
  linkcolor  = mywholecolor!\myshade!black,
  citecolor  = mywholecolor!\myshade!black,
  urlcolor   = mywholecolor!\myshade!black,
  colorlinks = true,
}

\title{Diff-MST: Differentiable Mixing Style Transfer}

\multauthor
{Soumya Sai Vanka$^{1\dag}$ \hspace{1cm} Christian Steinmetz$^{1\dag}$ \hspace{1cm} Jean-Baptiste Rolland$^2$} { \bfseries{Joshua Reiss$^1$ \hspace{1cm} Gy\"orgy Fazekas$^1$}\\
$^1$ Centre for Digital Music, Queen Mary University of London, UK\\
$^2$ Steinberg Media Technologies GmbH, Germany\\
{\tt\small s.s.vanka@qmul.ac.uk, c.j.steinmetz@qmul.ac.uk}
}

\def\authorname{S. Vanka, C. Steinmetz, J.-B. Rolland, J. Reiss, and G. Fazekas}

\begin{document}

\maketitle
\def\thefootnote{\textdagger}\footnotetext{These authors contributed equally to the work.}
\def\thefootnote{\arabic{footnote}}
\begin{abstract}
Mixing style transfer automates the generation of a multitrack mix for a given set of tracks by inferring production attributes from a reference song. However, existing systems for mixing style transfer are limited in that they often operate only on a fixed number of tracks, introduce artifacts, and produce mixes in an end-to-end fashion, without grounding in traditional audio effects, prohibiting interpretability and controllability. To overcome these challenges, we introduce \textbf{Diff-MST}, a framework comprising a differentiable mixing console, a transformer controller, and an audio production style loss function.
By inputting raw tracks and a reference song, our model estimates control parameters for audio effects within a differentiable mixing console, producing high-quality mixes and enabling post-hoc adjustments. Moreover, our architecture supports an arbitrary number of input tracks without source labelling, enabling real-world applications. We evaluate our model's performance against robust baselines and showcase the effectiveness of our approach, architectural design, tailored audio production style loss, and innovative training methodology for the given task. We provide code and listening examples online\footnote{\url{https://sai-soum.github.io/projects/diffmst/}}.
\end{abstract}

\section{Introduction}\label{sec:introduction}

\begin{figure}[t]
    \includegraphics[width=\linewidth,trim={0.9cm 0.0cm 0.6cm 0.0cm},clip]{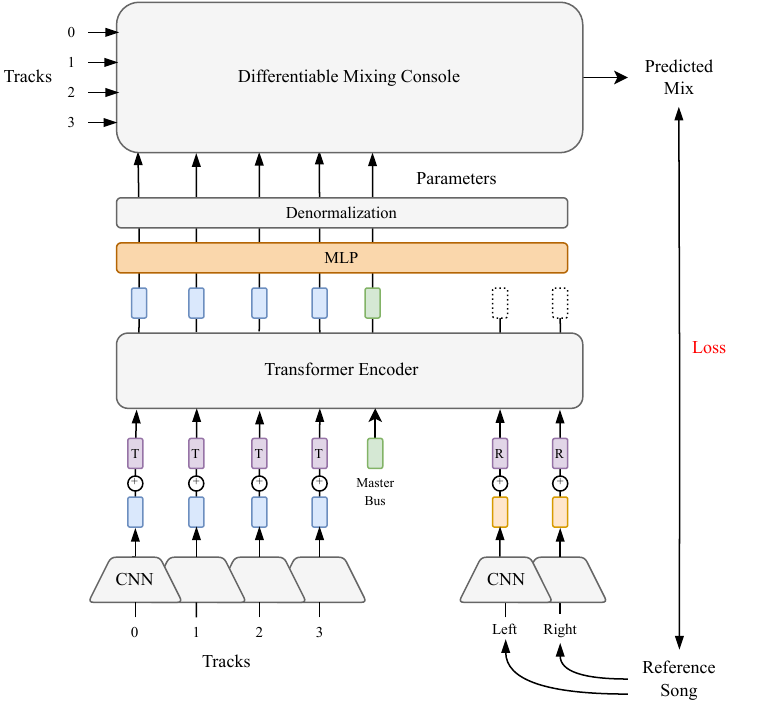}
    \caption{Diff-MST, a differentiable mixing style transfer framework featuring a differentiable multitrack mixing console, a transformer-based controller that estimates control parameters for this mixing console, and an audio production style loss function that measures the similarity between the estimated mix and reference mixes.}
    \label{fig:diff-mst}
\end{figure}
\begin{figure*}[t]
    \centering
    \includegraphics[width=0.75\linewidth,trim={0.9cm 0.4cm 1.4cm 0},clip]{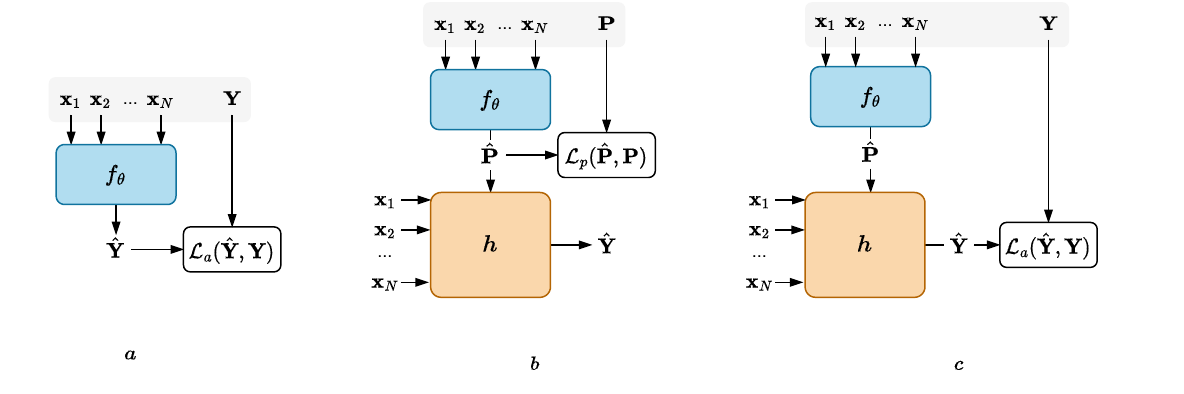}
    \caption{Formulations for deep learning-based automatic mixing systems \cite{steinmetz2022automix}. (a) Direct transformation (b) Parameter estimation on parameter loss (c) Parameter estimation on audio loss. Here, $x_i$ for $i \in [1, N]$ are the $N$ input tracks, $f_\theta$ is the transformation, $h$ is the dedicated mixing console, $Y$ and $\hat Y$ are the ground truth and predicted mix, $P$ and $\hat P$ are the ground truth and predicted control parameters and $L_a$ and $L_p$ are the audio and parameter loss respectively.}
    \label{fig:types-of-systems}
\end{figure*}

Music mixing involves technical and creative decisions that shape the emotive and sonic identity of a song~\cite{miller2016mixing}. The process involves creating a cohesive mix of the given tracks using audio effects to achieve balance, panorama, and aesthetic value~\cite{izhaki2017mixing}. Given the complexity of the task, mastering the task of mixing often requires many years of practice. To address this, several solutions have been proposed to provide assistance or automation~\cite{tenyearsai, steinmetz2022automix}. 
Automatic mixing systems have been designed using knowledge engineering~\cite{tom2019automatic, moffat2019machine}, machine learning, and more recently deep learning methods~\cite{martinez2021deep, steinmetz2021automatic, colonel2021reverse, steinmetz2020learning, martinez2022automatic}. 
Automatic mixing systems can be further subdivided into direct transformation systems and parameter estimation systems, as shown in Figure \ref{fig:types-of-systems}. Direct transformation systems operate on tracks and predict a mix directly, in an end-to-end fashion, with the loss calculated between the ground truth mix and the predicted mix. On the other hand, parameter estimation systems take input tracks and predict control parameters for a dedicated mixing console. In such systems, the loss can either be calculated on the predicted control parameters (parameter loss) based on the availability of ground truth, or on the predicted audio against the ground truth mix (audio loss). Parameter loss, calculated on the parameters, may not be optimal for multiparameter signal processing blocks since various combinations of parameters could potentially produce similar outcomes.
\cite{martinez2021deep, martinez2022automatic} utilizes a deep learning-based direct transformation system for mixing, while ~\cite{steinmetz2021automatic} employs a parameter estimation-based deep learning approach. However, many of these systems are constrained to a small number of input tracks or struggle to generalize effectively to real-world mixing scenarios. Furthermore, most of these approaches generate a mix without accounting for the desired sound and emotion. Due to the subjective nature of the task, an end-to-end approach without user control is less desirable in professional practice~\cite{sai2023adoption}.
\subsection{Mixing Style Transfer}
In professional practice, the audio engineer often uses reference songs and guidelines provided by the client to make mixing decisions~\cite{vanka2024role}. This encourages the development of automatic mixing systems that are aware of the intention of the mixing engineer. In our context, mixing style transfer refers to mixing in the style of given reference songs~\cite{vankamst2021}. This pertains to capturing the global sound, dynamics and spatialisation of the reference song. 
Recently, deep learning systems have been proposed for audio production style transfer. 
While some approaches have considered estimating the control parameters for audio effects~\cite{steinmetz2022style}, they are so far limited to controlling only a single or small set of effects with a singular input. 
Whereas ~\cite{koo2022music} have implemented an end-to-end style transfer system between two mixed songs which limits controllability and full raw tracks mixing. 
In this work, we introduce a novel deep learning-based approach to mixing multitrack audio material using a reference song, which utilises a differentiable mixing console to predict parameter values for gain, pan, 4-band equalization, compressor, and a master bus. Our proposed system is differentiable, interpretable and controllable, and can learn the mixing style from the given reference song. 
The contributions of this work can be summarised as follows:
\begin{enumerate}
    \item A framework for mixing style transfer that enables control of audio effects mapping the production style from a reference onto a set of input tracks. 
    \item A differentiable multitrack mixing console consisting of gain, parametric equalisation, dynamic range compression, stereo panning, and master bus processing using \texttt{dasp-pytorch}\footnote{\label{dasp}\url{https://github.com/csteinmetz1/dasp-pytorch/}}, which enables end-to-end training.
    \item Demonstration of the benefits of our system, including generalisation to an arbitrary number of input tracks, no requirement for labelling of inputs or enforcement of specific taxonomies, high-fidelity processing without artifacts, and greater efficiency.
\end{enumerate}
\begin{figure}
    \centering
    \includegraphics[width=0.8\linewidth,trim={0.0cm 0.3cm 0.0cm 0.2cm},clip]{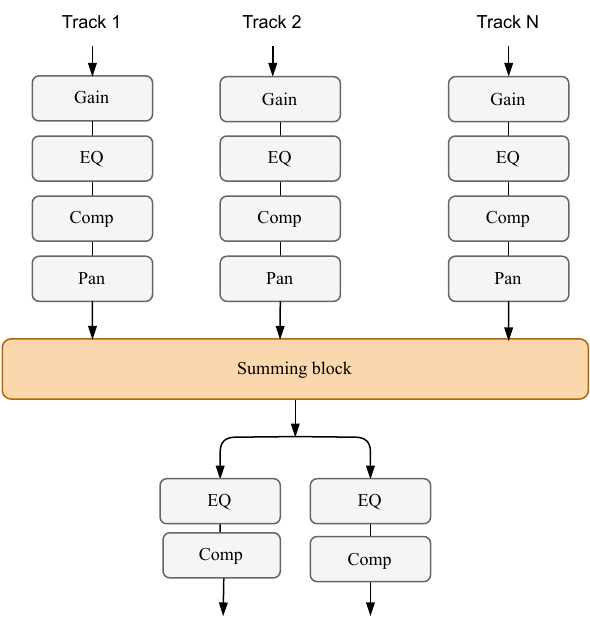}
    \caption{Differentiable Mixing console}
    \label{fig:DMC}
    \vspace{-0.3cm}
\end{figure}
\section{Method}
\subsection{Problem Formulation}
We can formulate the mixing style transfer task as follows. Let $T$ be a matrix of $N$ mono input raw tracks $\{t_1, t_2, t_3, \ldots, t_N\}$ and $M_r$ be the matrix of stereo reference mix containing two channels. A shared weight encoder $f_{\theta r}$ and $f_{\theta t}$ are employed to extract information from the reference and input tracks respectively. This information is then aggregated and fed into a transformer controller network comprising a transformer encoder and a multilayer perceptron (MLP) $g_\phi$. The primary task of this network is to estimate the parameter matrix $P$, which consists of $N$ parameter vectors $p$, each responsible for configuring the chain of audio effects for a respective track in $T$. Subsequently, the differentiable mixing console $h(T, P)$ processes the input tracks $T$ using the parameters $P$ to generate a predicted mix $M_p$ that mirrors the style of the reference mix $M_r$.

\begin{equation}
P = g_\phi(f_{\theta t}(T), f_{\theta r}(M_r))
\end{equation}
\begin{equation}
M_p = h(T, P)
\end{equation}
\subsection{Differentiable Mixing Style Transfer System}

We propose a differentiable mixing style transfer system (Diff-MST) that takes raw tracks and a reference mix as input and predicts mixing console parameters and a mix as output. 
As shown in Figure~\ref{fig:diff-mst}, our system employs two encoders, one to capture a representation of the input tracks and another to capture elements of the mixing style from the reference. A transformer-based controller network analyses representations from both encoders to predict the differentiable mixing console (DMC) parameters. The DMC generates a mix for the input tracks using the predicted parameters in the style of the given reference song. Given that our system oversees the operations of the DMC rather than directly predicting the mixed audio, we circumvent potential artefacts that may arise from neural audio generation techniques~\cite{pons2021upsampling, engel2020ddsp}. This also creates an opportunity for further fine-tuning and control by the user.
\subsection{Differentiable Mixing Console (DMC)}
The process of multitrack mixing involves applying a chain of audio effects, also known as a channel strip, on each channel of a mixing console.
The audio engineer may use these devices to reduce masking, ensure a balance between the sources, and address noise or bleed. Incorporating this prior knowledge of signal processing in the design of our mixing system, we propose an interpretable and controllable differentiable mixing console (DMC). 
Our console applies a chain of audio effects comprising gain, parametric equaliser (EQ), dynamic range compressor (DRC), and panning to each of the tracks to produce wet tracks. The sum of wet tracks is then sent to a master bus on which we insert stereo EQ and a DRC. This produces a mastered mix of the given tracks. We incorporate a master bus in our console as it is usual to use a mastered song as a reference in workflows. Therefore, having a master bus in the mixing console chain allows for easier optimisation of the system. To enable gradient descent and training in a deep learning framework, we require the mixing console to be differentiable. To achieve this, we use differentiable effects from the \texttt{dasp-pytorch}\textsuperscript{\ref{dasp}}. The pipeline of the DMC is presented in Figure~\ref{fig:DMC}.
\subsection{Spectrogram Encoder}
The encoder consists of a convolutional network based on the magnitude spectrum. It computes spectrograms by employing a short-time Fourier transform with a Hann window of size $N$ = 2048 and a hop size of $H$ = 512. The generated magnitude spectrogram is then processed through the convolutional layers. The resultant convolutional encodings are subsequently fed into a linear layer, producing a final embedding of size 512. The model includes separate shared-weight encoders: $f_{\theta r}$ for the reference mix and $f_{\theta t}$ for the input tracks. Each channel of stereo audio is treated as an individual track. Consequently, the stereo mix and any other stereo input tracks are loaded as separate tracks. Embeddings are computed by passing $T$ and $M_r$ through the encoder. 
\subsection{Transformer Controller}
The controller features a transformer encoder and a shared-weight MLP. The transformer encoder generates style-aware embeddings using self-attention across the output of the spectrogram encoder$f_{\theta r}$ and $f_{\theta t}$ and a master bus embedding which is learned during training. The MLP predicts the control parameters corresponding to the channel strip for each track, and the master bus embeddings are used to predict the master bus control parameters. A shared weight MLP is used to predict channel strip parameters for each channel. We generate the predicted mix $M_p$ by passing the control parameters through the DMC along with the tracks. This architecture enables our system to be invariant to the number of input tracks as shown in Figure~\ref{fig:diff-mst}. 
\subsection{Audio Production Style Loss}
\label{sec:audioloss}
The style of a mix can be broadly captured using features that describe its dynamics, spatialisation and spectral attributes ~\cite{vanka2024role}. We propose two different losses to train and optimise our models. \\

\noindent\textbf{Audio Feature (AF) loss}: This loss is composed of traditional Music Information Retrieval (MIR) audio feature transforms~\cite{man2014analysis}. The $T$ transforms include the root mean square (RMS) and crest factor (CF), stereo width (SW) and stereo imbalance (SI) and bark spectrum (BS) corresponding to the dynamics, spatialisation and spectral attributes respectively. We optimise our system by calculating the weighted average of the mean squared error on the audio features that minimises the distance between $M_p$ and $M_r$. We compute the audio feature transforms $T$ along with the weights $w$ as follows:
\begin{equation}
 \text{$T_1$}(\mathbf{x}) = \text{RMS}(\mathbf{x}) = \sqrt{\frac{1}{N}\sum_{i=1}^{N} x_i^2}\quad ; w_1 = 0.1
\end{equation}
\begin{equation}\text{$T_2$}(\mathbf{x}) =\text{CF}(\mathbf{x}) = 20 \log_{10}\left(\frac{\max(\lvert x_i \rvert)}{\text{RMS}(\mathbf{x})}\right)\quad; w_2 = 0.001
\end{equation}
\begin{equation}\text{$T_3$}(\mathbf{x}) =\text{BS}(\mathbf{x}) = \log(\mathbf{FB}\cdot \lvert \text{STFT}(\mathbf{x}) \rvert + \epsilon)\quad; w_3 = 0.1
\end{equation}
\begin{equation}\text{$T_4$}(\mathbf{x}) =\text{SW}(\mathbf{x}) = \frac{\frac{1}{N}\sum_{i=1}^{N}(x_{Li} - x_{Ri})^2}{\frac{1}{N}\sum_{i=1}^{N}(x_{Li} + x_{Ri})^2}\quad; w_4 =1.0
\end{equation}

\begin{equation}\text{$T_5$}(\mathbf{x}) =\text{SI}(\mathbf{x}) = \frac{\frac{1}{N}\sum_{i=1}^{N}x_{Ri}^2 - \frac{1}{N}\sum_{i=1}^{N}x_{Li}^2}{\frac{1}{N}\sum_{i=1}^{N}x_{Ri}^2 + \frac{1}{N}\sum_{i=1}^{N}x_{Li}^2}\quad; w_5 = 1.0
\end{equation}
\\
where N represents the sequence length, x is the input tensor, \textbf{FB} is the filterbank matrix, STFT(x) represents the short-time Fourier transform of x, and $\epsilon$ is a small constant of value $10^{-8}$ added for numerical stability. $x_{Li}$ and  $x_{Ri}$ represent the input tensor corresponding to the left and right channels, respectively. The net loss is computed as follows:
\begin{equation}
\text{Loss}(\mathbf{M_p}, \mathbf{M_r}) = \frac{1}{2} \sum_{i=1}^{2} \sum_{j=1}^{5} w_{j} \cdot \text{MSE}\left(\text{T}_{j}(\mathbf{M_p}_i), \text{T}_{j}(\mathbf{M_r}_i\right)
\end{equation}
where $w_j$ is the weight associated with $j^{th}$ transform $T_j$ and MSE corresponds to mean squared error. The weights for the transforms were determined through empirical testing to balance the scale of various losses.\\
\textbf{MRSTFT loss}: The multi-resolution short-time Fourier transform loss ~\cite{wang2019neural, steinmetz2020auraloss} is the sum of $L_1$ distance between STFT of ground truth and estimated waveforms measured in both log and linear domains at multiple resolutions, with window sizes $W \in [512, 2048,8192]$ and hop sizes $H =W/2$.
This is a full-reference metric meaning that the two input signals must contain the same content. 
\section{Experiment Design}
The task requires a dataset with multitrack audio, style reference, and the ground truth mix of the multitrack in the style of the reference for training. However, due to the lack of suitable datasets, we deploy a self-supervised training strategy to enable learning of the control of audio effects without labelled or paired training data. We achieve this by training our model under two different regimes which mainly vary in data generation and loss function. \\

\begin{figure}[t]
    \centering
    \includegraphics[scale = 0.85]{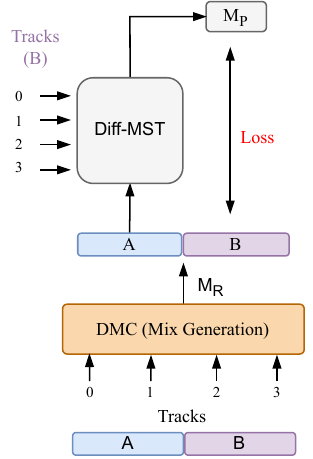}
    \caption{First training strategy from Section~\ref{sec:method_1}.}
    \label{fig:method_1}
    \vspace{-0.4cm}
\end{figure}

\noindent\textbf{Method 1}: We extend the data generation technique used in \cite{steinmetz2022style} to a multitrack scenario as shown in Figure~\ref{fig:method_1}. We first randomly sample a $t=10$\,s segment from input tracks and generate a random mix of these input tracks by using random DMC parameters. We then split the segment of the randomly mixed audio and the input tracks into two halves, namely, $M_{rA}$ and $M_{rB}$ and $T_A$ and $T_B$ of $t/2$\,s each, respectively. The model is input with $T_B$ as input tracks and $M_{rA}$ as the reference song. The predicted mix $M_p$ is compared against $M_{rB}$ as the ground truth for backpropagation and updating of weights. Using different sections of the same song for input tracks and reference song encourages the model to focus on the mixing style while being content-invariant. This method allows the use of MRSTFT loss for optimisation as we have the ground truth available. The predicted mix is loudness normalised to -16.0\,dBFS before computing the loss. 
\label{sec:method_1}
 \\

\noindent\textbf{Method 2}: We sample a random number of input tracks between 4-16 for song A from a multitrack dataset and use a pre-mixed real-world mix of song B from a dataset consisting of full songs as the reference. We train the model using AF loss mentioned in Section \ref{sec:audioloss} computed between $M_p$ and $M_r$. This method also allows us to train the model without the availability of a ground truth. 
Unlike Method 1, this approach exposes the system to training examples more similar to real-world scenarios where the input tracks and the reference song come from a different song.
However, due to random sampling, some input track and reference song combinations may not be realistic.
\label{sec:method_2}

\subsection{Datasets}
\noindent\textbf{Multitrack}: For both training methods, we utilise multitrack from MedleyDB ~\cite{bittner2014medleydb, bittner2016medleydb} and Cambridge.mt\footnote{\label{cmt}\url{https://cambridge-mt.com/}} which contains a total of 196 and 535 songs respectively, sampled at $f_s$ = 44100 Hz. For both datasets, we generate a train/test/validation split of 4:1:1. During training, songs are picked at random from the training split of both datasets. Thereafter, we randomly sample a section of the song as input tracks. We find a random offset for sampling multitrack by finding a section of the mix $x[i]$ that has mean energy above the threshold, $\frac{1}{N}\sum_{i=1}^{N}|x[i]|^{2} \geq 0.001$. During training, each channel corresponding to a stereo raw track is treated as a separate mono track. We check the mean energy of each track to avoid loading silent tracks. All input tracks are loudness normalised to -48.0 dBFS. \\

\noindent\textbf{Reference Songs}: For Method 1 we generate a random mix using random parameters and input tracks as mentioned in Section \ref{sec:method_1} and loudness normalise the random mix to -16 dBFS. 
For Method 2, we use real-world songs from MTG-Jamendo which contains more than 55k songs in MP3 format~\cite{bogdanov2019mtg}. We pick a random segment $y[i]$ of a random song from the dataset as a reference and check for mean energy above the threshold, $\frac{1}{N}\sum_{i=1}^{N}|xy[i]|^{2} \geq 0.001$. We loudness normalise the reference to -16 dbFS and load stereo information on separate channels.
\subsection{Training Details}
\label{sec:experiments}
Our model contains 190\,M trainable parameters, 76.5M corresponding to the track and mix encoder, and 37.9\,M for the transformer controller. We train five variations of our model differing in the number of tracks, methodology and loss function used. To remedy the bottleneck of reading multitrack audio data from disk, we load data into RAM every epoch from both the training and validation sets respectively.
The number of training steps per epoch is comprised of passing over these examples 20 times for training and 4 times for validation, sampling random examples at each step.
This provides a tradeoff between training speed and data diversity. 
We train all our models with a batch size of 2 and a learning rate of $10^{-5}$ with the \textit{Adam} optimiser. We accumulate gradients over 4 batches and use \verb|pytorch| for training. \\

\noindent\textbf{Diff-MST-MRSTFT}: We generate data using the method 1 described in Section~\ref{sec:method_1} and calculate MRSTFT loss for weight update and backpropogation. We train two variations of the model with a maximum of 8 tracks and 16 tracks as input, each for 1.16\,M steps. \\

\noindent\textbf{Diff-MST-MRSTFT+AF}: We fine-tune both versions of the pre-trained Diff-MST-MRSTFT using the synthetically generated data of method 1 in Section~\ref{sec:method_1} with AF loss described in Section \ref{sec:audioloss} for 20k steps. \\

\noindent\textbf{Diff-MST-AF}: We follow the training strategy mentioned in method 2 of Section~\ref{sec:method_2}  and use real-world songs as the reference. We train this model for 1.16\,M steps using the AF loss described in Section~\ref{sec:audioloss}. We train with a varying number of tracks with an upper limit of 16.
\subsection{Baselines}
\label{sec:baseline}
We compare the performance of our model against three baselines: an equal loudness mix (lowest anchor), the mix generated using the pre-trained mixing style transfer (MST) model by~\cite{koo2022music} (state-of-the-art), and two human mixes. We picked three songs from the Cambridge online multitrack repository belonging to the genres of electronic, pop, and metal for our main evaluation. Each of the songs contains between 12 and 22 input tracks. We selected references from popular songs. \\

\noindent\textbf{Equal Loudness}: We loudness normalise the tracks to -48.0\,dBFS and take the mean among the tracks to generate the mix which is then normalised. This generates a loudness-normalised sum of input tracks. We consider this system to be the lowest anchor as it does not consider any style information or mixing transformations.\\

\noindent\textbf{MST} ~\cite{koo2022music}: The method uses a pre-trained source separation model to generate stems from input and reference mix and perform stem-to-stem style transfer using a contrastive learning-based pre-trained audio effect encoder. The stems are mixed using a TCN-based model conditioned on style embeddings. Since the model performs a mix-to-mix transformation, we make use of the equal loudness mix of input tracks as the input to be transformed by the model. This allows us to extend the system to perform mixing style transfer for any number of input tracks. This puts the system at a disadvantage as it is trained to work for mix-to-mix scenarios where good-quality mixes are used as input, leading to better-quality extracted stems. 
\\

\noindent\textbf{Human Mixes}: We asked two audio engineers with professional practice to mix the three songs using the corresponding references. Each of them mixed all three songs until the end of the first chorus.
\section{Objective Evaluation}
We evaluate the performance of our model against three baselines listed in Section \ref{sec:baseline}. For the first evaluation, we compare the mixes generated by all five of our systems described in Section \ref{sec:experiments} and the baselines for three songs belonging to the genres of pop, electronic and metal. We manually picked the songs for the input tracks and the references for each of these cases. A 10-second section ranging between the middle of the first verse to the middle of the first chorus was used for evaluation in Table \ref{table:netAF}. We loudness normalise the reference mix to -16 dBFS and the predicted mix to -22 dBFS before predicting the metrics. \\
We report the average AF loss and individual weighted audio feature transforms from Section~\ref{sec:audioloss} for all three songs. Our Diff-MST system trained on real-world songs as reference using AF loss performs the best, closely followed by the MST~\cite{koo2022music}, human engineer mix, and the mix from our Diff-MST-MRSTFT+AF-16 system. \\
For the second evaluation, we compute average metrics across 100 randomly sampled examples with multitrack taken from the unseen set of Cambridge multitrack and reference songs from MUSEDB18~\cite{musdb18}. We compare the performance of our systems and the baselines MST~\cite{koo2022music} and the equal loudness system as shown in Table~\ref{tab:100compare}. We report individual weighted audio features from the AF loss along with average loss and Frèchet Audio distance (FAD)~\cite{kilgour2019frechet}. The FAD metric is employed to gauge the efficacy of music enhancement approaches or models by comparing the statistical properties of embeddings generated by their output to those of embeddings generated from a substantial collection of clean music. In this context, we analyze the distributions of real-world songs against the mixes generated by various systems using the VGGish model. Again, Diff-MST-AF-16 outperforms other approaches at capturing the dynamics, spatialisation and spectral attributes of the reference songs. 
\begin{table}[t!]
    \centering
    \footnotesize
    \setlength{\tabcolsep}{4pt}
    \begin{tabular}{lllllll}
        \toprule
        \textbf{Method}       &   \textbf{RMS} $\downarrow$   &   \textbf{CF}  $\downarrow$   &   \textbf{SW}  $\downarrow$   &   \textbf{SI}  $\downarrow$   &   \textbf{BS} $\downarrow$   &  \textbf{AF Loss} $\downarrow$  \\
        \midrule
        Equal Loudness   &3.11 & 0.51  & 3.16  & 0.21 &  33.3  &    33.389    \\
        MST~\cite{koo2022music}          & 3.15 & 0.45  & 4.64  & \textbf{0.13} & \textbf{0.09} &   \underline{0.185}    \\
        \midrule
        \textbf{Diff-MST} & & & & & & \\
        \scriptsize MRSTFT-8    & 3.63 &  1.44  & 1.97  & 4.29  & 0.17  &   0.379    \\
        \scriptsize MRSTFT-16   & 3.40  & 0.98  & 1.91  & 1.99  & 0.19  &   0.328    \\
        \scriptsize MRSTFT+AF-8  & 3.12 & 0.86  & 1.29  & 0.76 & 0.13  &   0.237    \\
        \scriptsize MRSTFT+AF-16  & 3.15 & 0.43  & \textbf{0.89} &  2.20  & 0.11  &   \underline{0.186}    \\
        \scriptsize AF-16     & \textbf{2.39} & \textbf{0.07} &  1.60  & 0.97 & 0.13  &   \textbf{0.168}    \\
        \midrule
        Human 1       & 3.02 &  0.26  & 2.05  & 0.46 & 0.17  &   0.218    \\
        Human 2       & 3.21 & 0.14  & 3.63  & 2.29  & 0.11  &    \underline{0.180}    \\
        \bottomrule
    \end{tabular}
    \caption{Average of metrics computed across the same section of three songs from three different genres. RMS is reported in e-04, CF in e-01, SW in e-02, and SI in e-02. We have provided audio examples as supplementary material.}
    \label{table:netAF}
\end{table}
\begin{table}[t]
    \setlength{\tabcolsep}{2.5pt}
    \centering
    \footnotesize
    \begin{tabular}{l l l l l l l l}
        \toprule
        \textbf{Method} & \textbf{RMS}  $\downarrow$ & \textbf{CF} $\downarrow$ & \textbf{SW} $\downarrow$ & \textbf{SI} $\downarrow$ & \textbf{BS} $\downarrow$ & \textbf{AF loss} $\downarrow$ &  \textbf{FAD} $\downarrow$ \\ \midrule 
        Equal Loudness & \textbf{2.31e-04} & 2.11 & 6.03 & 1.41 & 32.7 & \text{6.55e+00} & 17.6 \\ 
        MST~\cite{koo2022music} & \text{4.07e-04} & 1.72 & 5.84 & 0.89 & 0.31 & \underline{\text{7.85e-02}} & 17.9 \\ 
        \midrule
        \textbf{Diff-MST} & & & & & & \\
        \scriptsize \text{MRSTFT-8} & \text{3.08e+06} & 3.91 & 4.55 & 3.38 & 7.06 & \text{6.15e+05} & 51.3\\ 
        \scriptsize \text{MRSTFT-16} & \text{2.23e+03} & 4.07 & 5.00 & 1.97 & 1.81 & 4.47e+02 & 65.9\\ 
        \scriptsize \text{MRSTFT+AF-8} & \text{2.00e+05} & 1.79 & 4.58 & 2.86 & 6.89 & \text{4.00e+04} & 48.3 \\ 
        \scriptsize \text{MRSTFT+AF-16} & \text{2.46e+00} & 1.14 & \textbf{4.29} & 3.44 & 0.92 & 6.92e-01 &51.1 \\ 
        \scriptsize \text{AF-16} & \text{4.24e-04} & \textbf{0.67} & 4.78 & \textbf{0.22} & \textbf{0.11} & \textbf{3.26e-02} &\textbf{15.1} \\ 
        \bottomrule
    \end{tabular}
    \caption{Average of metrics using unseen tracks from Cambridge dataset and mixes from MUSDB18 ~\cite{musdb18}. CF in e-02, SW in e-02, SI in e-02.}
    \label{tab:100compare}
\end{table}
\section{Discussion}
Overall, the results indicate the effectiveness of our approach, architecture choice, custom audio production style loss, and novel training regime for the task. The reported metrics for both evaluations show improved performance when trained on a larger number of tracks. Furthermore, we also see that the systems trained or fine-tuned using AF loss generally perform better than those trained with MRSTFT loss, specifically in improving the spatialisation and dynamics of the mixes, thus showing the efficacy of our hand-crafted audio feature-based loss function. \\
The significant difference in the Bark spectrum values between the equal loudness and our system's mixes suggests that mixes generated using our system have undergone significant spectral processing, resulting in an increased spectral similarity between the reference song and the predicted mix. The metrics indicate inferior performance for the Diff-MST-MRSTFT-8/16 model compared to all our proposed models. This may be attributed to the training data, which is generated using random mixing console parameters, often resulting in mixes that sound unrealistic. However, fine-tuning with AF loss during the last steps notably enhances performance. This improvement could be attributed to AF loss compelling the model to enhance dynamics and spatialization, as evidenced by the reported metrics. We observe a notable enhancement in performance through training on real-world songs, underscoring the significance of high-quality real-world data. \\
Although the system demonstrates promising outcomes, it is not without its limitations. While we note higher metric values for certain features on the human mixes, this can be explained by the fact that human engineers often strive to capture the overall essence of the reference song. However, they may also incorporate creative elements leading to spatialization and dynamics that diverge significantly from the reference. Our metrics serve to quantify the similarity between the reference song and the predicted mix, which is suitable for the task at hand but may fall short in assessing the creative or unconventional decisions made by human engineers during the mixing process. 
Additionally, while FAD indicates the predicted audio quality, it might not capture the intricate nuances involved in the mixing process, such as frequency masking and achieving balance and spatialization.\\
Moreover, we noticed a decline in the system's mixing capabilities as the number of input tracks increased beyond what it was trained on. Additionally, our mixing console lacks a crucial reverb module essential for comprehensive mixing tasks. Determining the optimal method for processing the entire song poses a challenge, as inferring over the entire song length may result in overly sparse embeddings. 
Our current system also falls short in modelling mixing context in all possible senses as discussed in ~\cite{lefford2021context}. However, we address this challenge by incorporating a reference input, typically selected by the mixing engineer or client. The reference song serves as a proxy for some of the contextual information that engineers typically rely on when making mixing decisions. Lastly, while real-world mixing often entails dynamic adjustments to effect parameters over the course of a song, our system is presently constrained to static mixing configurations.
\section{Conclusion}
In this work, we proposed a framework for mixing style transfer for multitrack music using a differentiable mixing console. Our system is rooted in strong inductive bias, taking inspiration from real-world mixing consoles and channel strips and predicts control parameters for these signal processing blocks allowing interpretability and controllability. Our system supports inputting any number of raw tracks, without source labelling. Furthermore, we circumvent possibilities for audio degradation and artifacts with our design choice for a parameter estimation-based system. Objective evaluations demonstrate that our Diff-MST-MRSTFT+AF-16 system surpasses all baseline methods. The reported metrics give us an insight into the impact of architectural and training design choices. We show that training on a larger number of input tracks improves the performance substantially while running inference on real-world examples that generally contain a larger number of input tracks. We also demonstrate the benefits of training on real-world quality audio examples.\\
While our research has produced promising results based on objective metrics, it is important to acknowledge our evaluation's constraints, as we have not conducted subjective assessments via listening tests. While objective metrics offer valuable insights into the model's performance, integrating subjective evaluations would provide a more comprehensive understanding of its efficacy in practical applications. Future work includes conducting an extensive subjective evaluation alongside assessing the usability of a prototype of the system that is integrated into the real-world workflow in the digital audio workstation (DAW). Further, work towards developing a robust understanding and objective metrics for mix similarity and mixing style is imperative for enhancing these systems.

\section{Acknowledgments}
We express our sincere gratitude to the ISMIR reviewers for providing valuable feedback on our work.
Further, we extend our thanks to Steinberg's research and development team for their unwavering support and honest feedback throughout this project.

This work is funded and supported by UK Research and Innovation [grant number EP/S022694/1] and Steinberg Media Technologies GmbH under the AI and Music Centre for Doctoral Training (AIM-CDT) at the Centre for Digital Music, Queen Mary University of London, London, UK.  

\section{Ethical Statement}
We utilized open-source multitrack data from MedleyDB~\cite{bittner2014medleydb, bittner2016medleydb} and the web forum Cambridge.mt\textsuperscript{\ref{cmt}} as well as full songs from MTG-Jamendo~\cite{bogdanov2019mtg} to train our models. MedleyDB and MTG-Jamendo are available under the licenses CC-BY-NC-SA and Apache 2.0, respectively. Cambridge.mt is an educational web platform managed by Mike Senior, a professional mixing engineer, where artists and professional engineers consensually share audio files for multitracks and corresponding mixes. The terms and conditions permit educational and non-commercial research usage. \\
The design of our system integrates user-centric principles and has been built upon extensive qualitative research involving professional engineers~\cite{vanka2024role}. Moreover, the design of this system is grounded in traditional mixing methods and expert knowledge, incorporating context and ensuring controllability and interpretability. In professional environments, our system can provide technical assistance for mixing, allowing more time for creative expression.  Additionally, our system aims to make music mixing more accessible for beginners and non-specialists, promoting the democratisation of music production.  The system can be used as a tool for learning basic mixing skills using the reference method. This system has the potential to support musicians and bands to create and distribute their music affordably, resulting in diverse representation within the music industry. However, there are potential drawbacks. Automated mixing systems might reduce the need for professional audio engineers in budget productions, impacting their job opportunities. Moreover, the widespread use of these tools may lead to homogenization in music production, resulting in algorithmically driven mixes overshadowing unique stylistic traits.

\bibliography{ISMIRtemplate}

%
%
%
%
%

\end{document}